\documentclass[aps,prl,twocolumn,groupedaddress,showpacs,nofootinbib]{revtex4-1}
\usepackage{graphicx}
\usepackage{dcolumn}
\usepackage{bm}
\usepackage{hyperref}
\usepackage{upgreek}
\usepackage{color}
\usepackage{soul}
\usepackage{units}

\newcommand{\bs}[1]{\boldsymbol{#1}}

\begin{document}
\title{Scaling of the Quantum Anomalous Hall Effect as an Indicator of Axion Electrodynamics}

\author{S. Grauer}
\affiliation{Faculty for Physics and Astronomy (EP3 and TP1),
Universit\"at W\"urzburg, Am Hubland, D-97074, W\"urzburg, Germany}
\author{K. M. Fijalkowski}
\affiliation{Faculty for Physics and Astronomy (EP3 and TP1),
Universit\"at W\"urzburg, Am Hubland, D-97074, W\"urzburg, Germany}
\author{S. Schreyeck}
\affiliation{Faculty for Physics and Astronomy (EP3 and TP1),
Universit\"at W\"urzburg, Am Hubland, D-97074, W\"urzburg, Germany}
\author{M. Winnerlein}
\affiliation{Faculty for Physics and Astronomy (EP3 and TP1),
Universit\"at W\"urzburg, Am Hubland, D-97074, W\"urzburg, Germany}
\author{K. Brunner}
\affiliation{Faculty for Physics and Astronomy (EP3 and TP1),
Universit\"at W\"urzburg, Am Hubland, D-97074, W\"urzburg, Germany}
\author{R. Thomale}
\author{C. Gould}
\affiliation{Faculty for Physics and Astronomy (EP3 and TP1),
Universit\"at W\"urzburg, Am Hubland, D-97074, W\"urzburg, Germany}
\author{L. W. Molenkamp}
\affiliation{Faculty for Physics and Astronomy (EP3 and TP1),
Universit\"at W\"urzburg, Am Hubland, D-97074, W\"urzburg, Germany}

\date{\today}

\begin{abstract}
We report on the scaling behavior of V-doped (Bi,Sb)$_2$Te$_3$ samples in the quantum anomalous Hall regime for samples of various thickness. While previous quantum anomalous Hall measurements showed the same scaling as expected from a two-dimensional integer quantum Hall state, we observe a dimensional crossover to three spatial dimensions as a function of layer thickness. In the limit of a sufficiently thick layer, we find scaling behavior matching the flow diagram of two parallel conducting topological surface states of a three-dimensional topological insulator each featuring a fractional shift of  $\frac{1}{2} e^2/h$ in the flow diagram Hall conductivity, while we recover the expected integer quantum Hall behavior for thinner layers. 
This constitutes the observation of a distinct type of quantum anomalous Hall effect, resulting from $\frac{1}{2} e^2/h$ Hall conductance quantization of three-dimensional topological insulator surface states, in an experiment which does not require decomposition of signal to separate the contribution of two surfaces. This provides a possible experimental link between quantum Hall physics and axion electrodynamics.

\end{abstract}
\maketitle

{\it Introduction.} After several theoretical proposals to accomplish the quantum anomalous Hall effect (QAHE)~\cite{Onoda2003,Liu2008,Yu2010,Qiao2010,Nomura2011}, it has recently been observed in magnetically doped (Bi,Sb)$_2$Te$_3$~\cite {Chang2013,Checkelsky2014,Kou2014,Bestwick2015,Chang2015a,Chang2015b,Feng2015,Kandala2015,Grauer2015}. The exact quantization and edge channel transport have been verified. Theory has suggested the system may be a suitable host for the realization of Majorana bound states~\cite{Qi2010} by combining superconductivity with the QAHE, and exploring quantized signatures of the three dimensional topological magneto-electric effect~\cite{Nomura2011,Wang2015,PhysRevB.92.085113}.

The QAHE has thus far been observed in thin layers with thicknesses in the range of 4 to 10 nm which are typically described as thin films. The onset of a gap stemming from the hybridization of the top and bottom surface state is predicted to be in the range of 6 nm~\cite{Liu2010,Lu2010}, which is also the thickness at which opening of the gap is observed in ARPES measurements on Bi$_2$Se$_3$ by Zhang \textit{et al.}~\cite{ZhangY2010}, and also in layers grown by our group~\cite{Landolt2014}. While many of the QAHE layers reported on so far are thinner than 6 nm, some are not. Also, layer roughness, resulting in part from the existence of rotation twins~\cite{Tarakina}, make it difficult to define an exact layer thickness. It is therefore not clear if the layers should be regarded as magnetic 2D or 3D topological insulators (TI). 

The mechanism invoked to explain the QAHE in its initial manifestation~\cite{Chang2013} is applicable to a 2D system~\cite{Yu2010}, where the inversion of one spin species is lifted by exchange interaction. The theoretical perspective on the QAHE at a 3D TI surface, however, is different, and ties to the axion term~\cite{Wilczek} characterizing the electrodynamic response of a 3D TI bulk~\cite{PhysRevB.78.195424}. This has recently been investigated through Faraday and Kerr rotation spectroscopy~\cite{armitage16,dziom16,Okada}, where one observes the joint effect of both topologically non-trivial surfaces and has no measure to decompose the signal into individual surface contributions. Due to the effect of the axionic action $S_\theta=\frac{ \theta \alpha}{4\pi} \int \bs{E} \cdot \bs{B} \; d^3 x dt $, where $\alpha$ is the fine structure constant and $\theta$=1 in the TI bulk up to its boundary, the single Dirac cone surface state does not violate gauge symmetry upon minimal coupling to an electromagnetic field ~\cite{PhysRevLett.51.2077,PhysRevLett.52.18}. As the magnetic dopants induce a gap and magnetic disorder might act to further localize the Dirac surface density of states, a half-integer contribution to the Hall conductivity $\sigma_{xy}=\pm \frac{1}{2}e^2/h$ is still expected to be observable as long as the Fermi level resides in the 3D TI bulk gap~\cite{Nomura2011}. The surface state of a 3D TI exhibiting the QAHE could thus be regarded as an ``axion insulator''.

In this letter, we report on our measurements on V$_y$(Bi$_{1-x}$Sb$_x$)$_{2-y}$Te$_3$ layers with thicknesses around $9 \text{ nm}$ which match the predicted flow diagram of two parallel topological surface states, providing experimental signature of a QAHE quantized in units of $\sigma_{xy}=\pm \frac{1}{2}e^2/h$ on each of the two surfaces, e.g. top and bottom, of a magnetic 3D TI slab. This effect is best evidenced by examining flow diagrams describing the relation of the longitudinal $\sigma_{xx}$ to the transversal $\sigma_{xy}$ conductivities during the transition from one filling factor to another. A flow diagram of the QAHE was first reported by Checkelsky \textit{et al.}~\cite{Checkelsky2014} and resembles that of the integer quantum Hall effect (iQHE)~\cite{Hilke1999} where the relation between $\sigma_{xx}$ and $\sigma_{xy}$ follows semicircles centered on $(\sigma_{xy}, \sigma_{xx}) = (\frac{1}{2} \text{ e}^2/\text{h}, 0)$ and $(-\frac{1}{2} \text{ e}^2/\text{h}, 0)$, going through the points $(\text{e}^2/\text{h},0)$, $(0,0)$ and $(- \text{e}^2/\text{h},0)$. Both components of the conductance going to zero during the transition indicate a complete breakdown of the edge channel transport.
By extracting the values of $\sigma_{xx}$ and $\sigma_{xy}$ from published measurements of the external magnetic field dependence to determine their scaling diagram, one finds that most subsequent observations of the QAHE show the same behavior. This can be directly seen by the peak value of the longitudinal resistivity $\rho_{xx} > \text{h}/\text{e}^2$ at $\rho_{xy} = 0$. Examples of such a high longitudinal resistivity are measurements with $\rho_{xx}=2.2 \text{ h}/\text{e}^2$  in Fig.~2 of~\cite{Chang2013}, $\rho_{xx}=2.1 \text{ h}/\text{e}^2$  in Fig.~2 of~\cite{Checkelsky2014},  $\rho_{xx}= 1.7 \text{ h}/\text{e}^2$  in Fig.~1 of~\cite{Chang2015a}, $\rho_{xx} =1.6 \text{ h}/\text{e}^2$  in Fig.~4 of~\cite{Chang2015b}, or even as high as $\rho_{xx}=34 \text{ h}/\text{e}^2$  in Fig.~1 of~\cite{Feng2015}. The lowest peak values of the longitudinal resistivity is seen in Fig.~1 of~\cite{Bestwick2015} with $\rho_{xx}=1.1 \text{ h}/\text{e}^2$ which, with a thickness of 10 nm, is measured on the thickest layer of the listed experiments. While our control measurements on thin layers show this same scaling behavior, our slightly thicker samples show the transition to the very different scaling behavior consistent with that predicted for an axion insulator. 

{\it Magnetic 3D TI layers.} Our V$_y$(Bi$_{1-x}$Sb$_x$)$_{2-y}$Te$_3$ layers are grown by molecular beam epitaxy on Si(111) and InP(111) substrates and capped \textit{in situ} with a 10 nm layer of Te as protection against the environment. The V content $y$ was determined from the growth rate of pure V to be $y \approx 0.1$ for all layers, the varying Sb content $x$ is determined by X-ray diffraction (XRD) measurements of the lateral lattice constant $a$ and the layer thicknesses are obtained from  X-ray reflection (XRR) measurements~\cite{Winnerlein2016}. After growth the layers are patterned into Hall bar devices with a top gate and AuGe contacts using standard optical lithography (see supplementary for more details). All presented transport measurements were conducted at base (nominally 25 mK) temperature (unless specified otherwise) in a dilution refrigerator equipped with a high field magnet. Hall and longitudinal resistances were measured using standard high-precision, low-frequency ac techniques (2-14 Hz). In all cases, the dependence of the longitudinal ($\rho_{xx}$) and Hall resistivity ($\rho_{xy}$) on the gate voltage was first measured in the absence of magnetic field, and all magneto-resistivity data presented in this letter is then taken for the gate voltage condition which gives the maximum value of $\rho_{xy}$ (unless specified otherwise).

Figure 1 shows a measurement of the longitudinal and Hall resistivity of several samples in an external magnetic field: Five 9 nm thick layers with varying composition V$_{0.1}$(Bi$_{1-x}$Sb$_x$)$_{1.9}$Te$_3$, a 8 nm thick V$_{0.1}$(Bi$_{0.22}$Sb$_{0.78}$)$_{1.9}$Te$_3$ layer and a 6 nm thick V$_{0.1}$(Bi$_{0.21}$Sb$_{0.79}$)$_{1.9}$Te$_3$ layer. The $\rho_{xy}$ data of Fig.~1a shows that all layers are in the QAH regime and at least close to quantization, with $\rho_{xy}(H=0) > 0.8 \text{ h}/\text{e}^2$. The longitudinal resistivity shown in Fig.~1b on the other hand shows results that can be classified in two distinct categories. For all layers with $d \approx 9 \text{ nm}$ the peak value of $\rho_{xx}$ is less than $\text{h}/\text{e}^2$, while the layers with $d < 9 \text{ nm}$ in contrast show a peak value which is higher than $\text{h}/\text{e}^2$, as common in the literature.
\begin{figure}%
\includegraphics[width=\columnwidth]{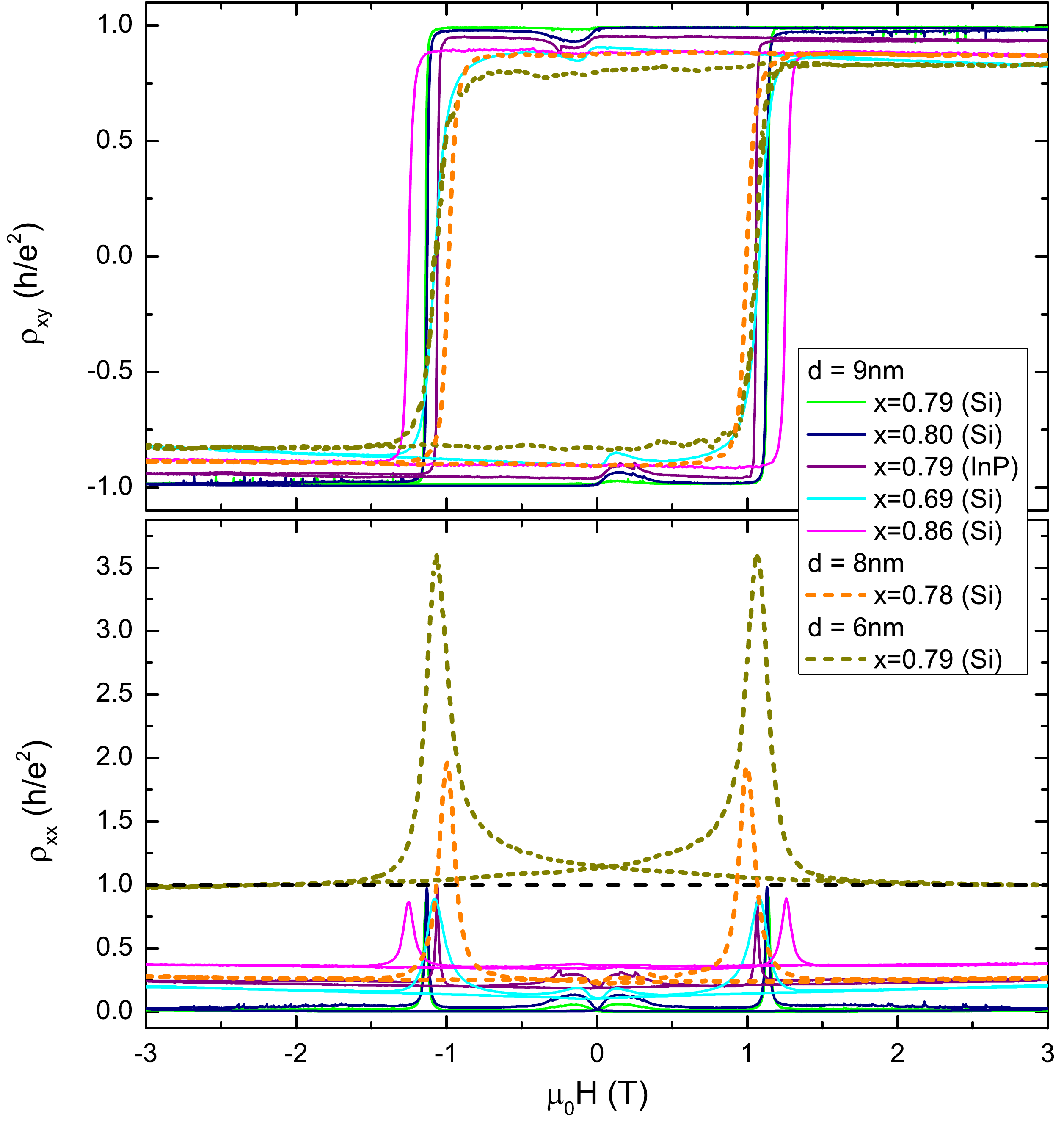}%
\caption{$\rho_{xy}$ (a) and $\rho_{xx}$ (b) of several V$_{0.1}$(Bi$_{1-x}$Sb$_x$)$_{1.9}$Te$_3$ layers as a function of the external magnetic field. The varied parameters are the Sb content $x$, the layer thickness $d$ and substrate type. The data acquired on the five layers with $d \approx 9 \text{ nm}$ is displayed as solid lines. Four of them are grown on Si(111) with $x$ values of: 0.69 (cyan), 0.79 (green), 0.80 (navy), 0.86 (magenta). One is grown on InP(111) with $x=0.79$ (purple). The measurements of two thinner layers with $d = 8 \text{ nm}$ and $x=0.78$ (orange) and $d = 6 \text{ nm}$ and $x=0.79$ (dary yellow), grown on Si(111), are displayed as dashed lines.}%
\label{fig:ExtFieldDep}%
\end{figure}

Flow diagrams of these measurements for the $d \approx 9 \text{ nm}$ layers are shown in Fig.~2. These diagrams show the scaling of $\sigma_{xx}$ to $\sigma_{xy}$ as an external parameter is used to turn the system between plateaus. Any external parameter which affects the plateau transition can be used. In this paper we choose to use external magnetic field as a parameter \cite{Kou2015,Hilke1998,Hilke1999} as it allows us to access a larger area of phase space while at the same time protecting the insulation nature of the bulk. Scaling diagrams using parameters such as temperature and gate voltage are shown in the supplemental information. The transition does not occur via the insulating state at (0,0), but rather different behavior is observed. Instead of following the above described flow diagram of the iQHE (black dashed lines), the data follows a semicircle centered on the origin with a radius of $\text{e}^2/\text{h}$ (red dashed line).

\begin{figure}%
\includegraphics[width=\columnwidth]{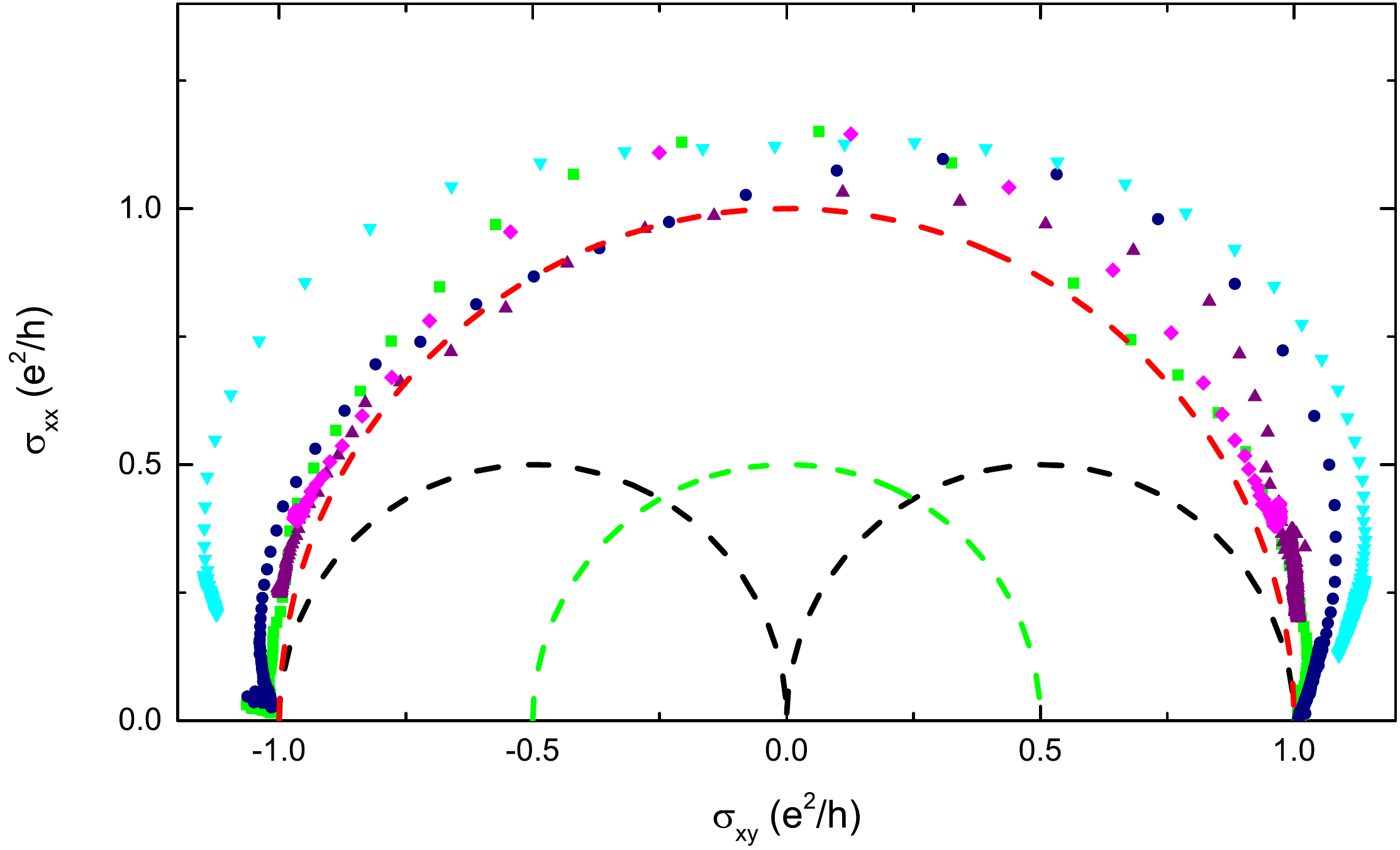}%
\caption{Flow diagram mapping $\sigma_{xx}$ to $\sigma_{xy}$ of layers with $d \approx 9 \text{ nm}$ from the external magnetic field measurement (see Fig.~1) is shown. The black dashed line represents the flow diagram of the iQHE, the green dashed line the predicted flow diagram of a single topological surface state and the red dashed line two parallel conducting topological surface states. The color code of the measurement data is the same as in Fig.~1.}%
\label{fig:ScalingDiffSb}%
\end{figure}

Such a semicircle centered on the origin is exactly what is predicted for a topological surface state in a magnetic 3D TI. Nomura and Nagaosa studied the transition of half-integer quantized states from $\sigma _{xy} = \frac{1}{2} \text{ e}^2/\text{h}$ to $\sigma _{xy} = -\frac{1}{2} \text{ e}^2/\text{h}$ on a single surface of a TI~\cite{Nomura2011}, and found a scaling behavior shown by the green dashed line in \mbox{Fig.~2}; a semicircle which connects the two points ($\frac{1}{2} \text{ e}^2/\text{h}$,0) and ($-\frac{1}{2} \text{ e}^2/\text{h}$,0) without transitioning through the insulating state. For two parallel conducting surface states, e.g. top and bottom surface of the layer, the conductivities add and the resulting scaling behavior is represented by the red dashed line in \mbox{Fig.~2}, consistent with our observed behavior. It is worth noting that the key distinction between the iQHE and the axionic behavior is the position of the center of the semicircle in the flow diagram, which is at $\sigma_{xy} = \frac{1}{2} \text{ e}^2/\text{h}$ for the former case, and shifted back by $\frac{1}{2} \text{ e}^2/\text{h}$ in the latter. This shift from finite value to zero does not depend on the number of surfaces being considered and thus is robust evidence of a fundamentally different scaling, which is consistent with an axionic term acting on a single topological surface state.  

\begin{figure}[t]%
\includegraphics[width=\columnwidth]{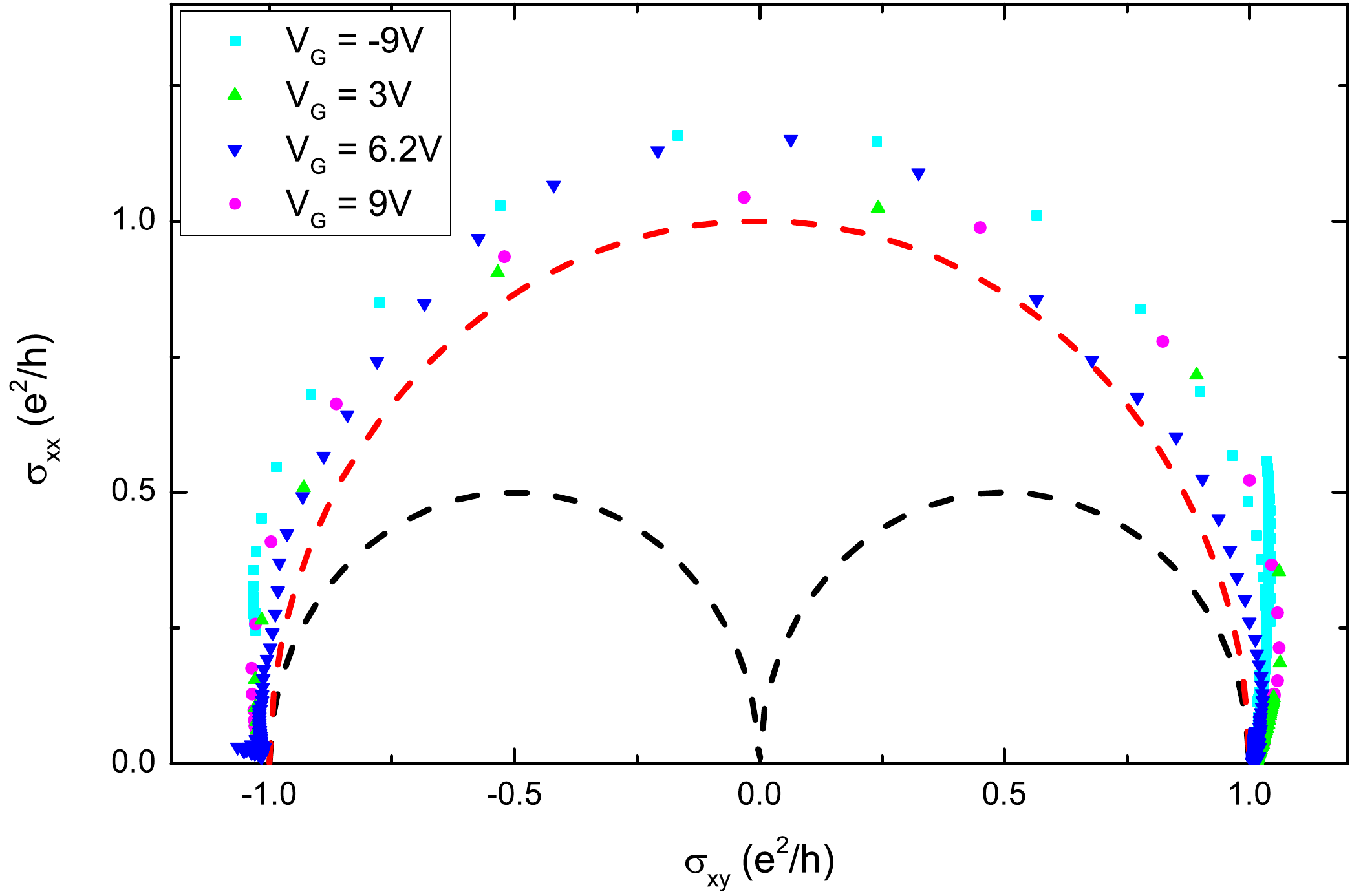}%
\caption{Flow diagram mapping $\sigma_{xx}$ to $\sigma_{xy}$ for 4 different fixed gate voltage values $V_G$ = -9 V (cyan), $V_G$ = 3 V (green), $V_G$ = 6.2 V (blue), and $V_G$ = 9 V (magenta), for a representative 9 nm thick V$_{0.1}$(Bi$_{0.21}$Sb$_{0.79}$)$_{1.9}$Te$_3$ layer exhibiting perfectly quantized transport (green in Fig.~1, Fig.~2, and Fig.~5). The black dashed line represents the flow diagram of the iQHE and the red dashed line two parallel conducting topological surface states.}%
\label{fig:ScalingDiffgate}%
\end{figure}

\begin{figure}[t]%
\includegraphics[width=\columnwidth]{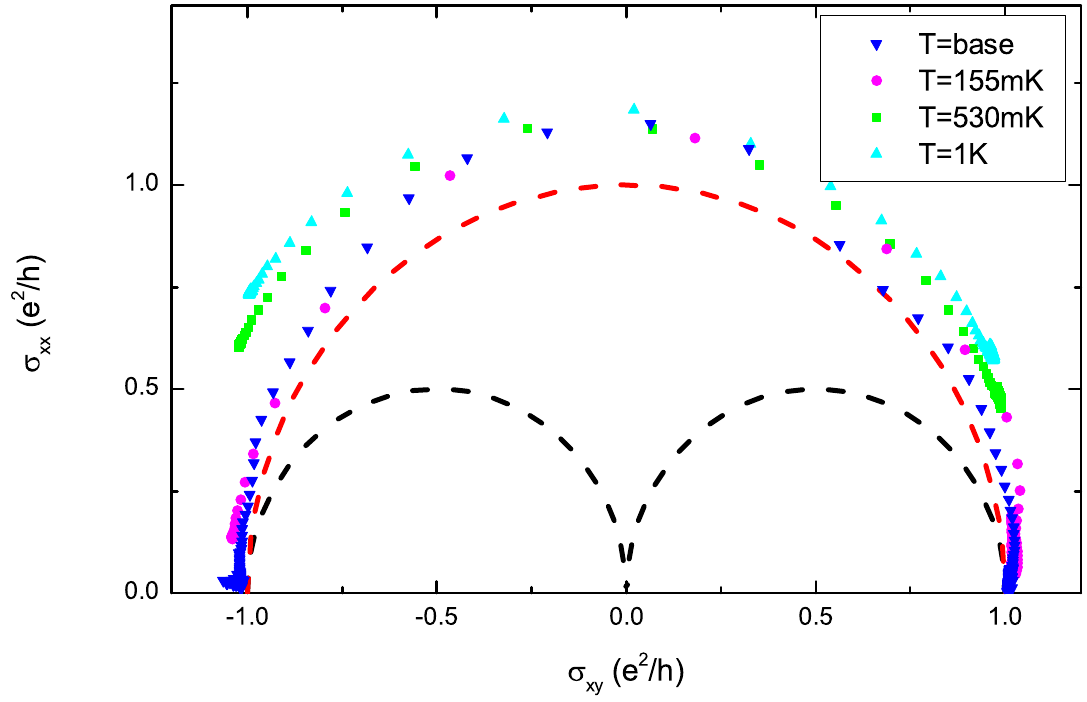}%
\caption{Flow diagram mapping $\sigma_{xx}$ to $\sigma_{xy}$ for 4 different fixed temperature values $T$ = 155 mK (magenta), $T$ = 530 mK (green), $T$ = 1 K (cyan), and base temperature of the dilution refrigerator (blue) for a representative 9 nm thick V$_{0.1}$(Bi$_{0.21}$Sb$_{0.79}$)$_{1.9}$Te$_3$ layer exhibiting perfectly quantized transport (green in Fig.~1, Fig.~2, and Fig.~5). The black dashed line represents the flow diagram of the iQHE and the red dashed line two parallel conducting topological surface states.}%
\label{fig:ScalingDiffT}%
\end{figure}

{\it Robustness of the effect.} To further confirm the robustness of the axionic scaling, we present flow diagram analysis for our representative perfectly quantized 9 nm thick V$_{0.1}$(Bi$_{0.21}$Sb$_{0.79}$)$_{1.9}$Te$_3$ layer \cite{Grauer2015}, for a wide range of fixed gate voltage values in Fig.~3, and for various fixed temperature values in Fig.~4. As expected, so long as the Fermi level resides in the bulk bandgap, change in carrier concentration does not affect the scaling properties of such a system. Similar reproducible behavior characterizes the magnetic field scans at elevated temperatures, as clearly seen in Fig.~4. Magneto-resistivity measurements at temperatures ranging from base temperature of the dilution refrigerator up to 1 K, reveal nearly constant longitudinal resistivity peak values, while high field values increase from 0 to nearly 0.5 $\text{h}/\text{e}^2$, as seen in supplementary Fig.~4. The same conclusion applies to magneto-resistivity scans for different gate voltage values visible in supplementary Fig.~3, where the $\rho_{xx}$ peak value again proves to be nearly carrier concentration independent and close to the quantized value $\text{h}/\text{e}^2$.

{\it Sensitivity to layer thickness.} We first note that because (Bi,Sb)$_{2}$Te$_3$ grows with rotational twins, no perfectly flat layer in this material system has ever been achieved, and the typical surface roughness is not negligible compared to layer thicknesses. As such layer thicknesses reported by various groups using methodologies, which are effected differently by this roughness are difficult to compare. The values of the layer thicknesses given here should therefore be viewed as a way to reliably compare the relative thicknesses of our samples, but will not necessary compare in absolute values to those reported by other groups. Having said that, to confirm that thickness is indeed the key parameter distinguishing between iQHE and axionic scaling, we reproduce the scaling curve of a 9 nm thick layer exhibiting perfectly quantized transport, together with that of an 8 nm and a 6 nm thick layer of similar composition in Fig.~5. Already for the 8 nm thick layer, a clear deviation  from the 3D TI axionic scaling behavior is seen in the form of a dip at $\sigma_{xy} = 0$. This feature strengthens for the 6 nm thick layer, which qualitatively follows the iQHE flow diagram represented by the black dashed line. Although both the 8 nm and 6 nm layers do not accurately quantize, they are in the QAHE regime and show a transition in scaling behavior, from the axion-like scaling expected in the 3D case to the previously reported iQHE scaling, highlighting the existence of two distinct regimes of QAHE. Unfortunately layers of even greater thickness cannot be studied as bulk contributions begin to significantly influence the observed transport as shown in supplementary Fig.~2.
   
\begin{figure}[t]%
\includegraphics[width=\columnwidth]{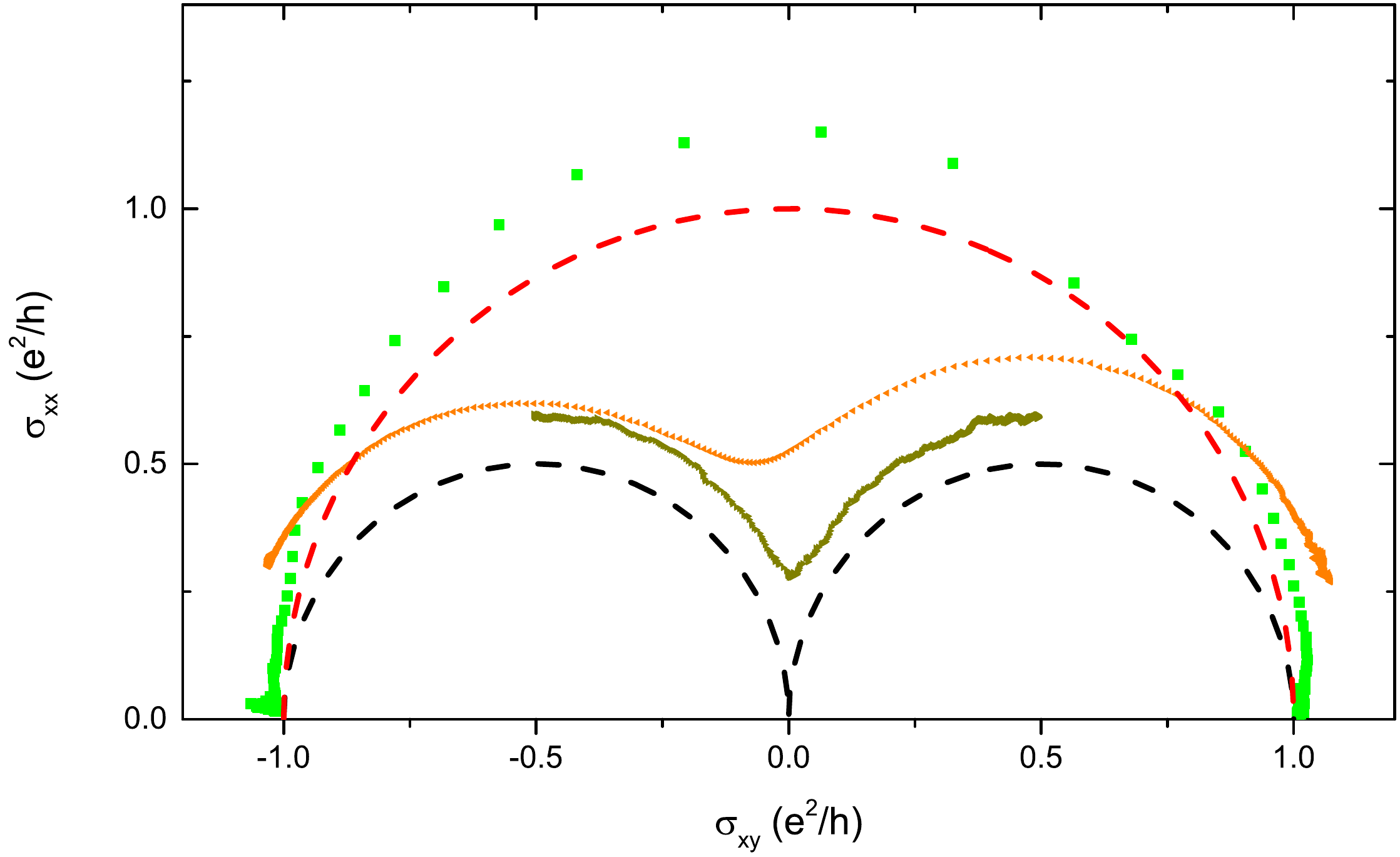}%
\caption{Flow diagram mapping $\sigma_{xx}$ to $\sigma_{xy}$ for 3 different layers with $d = 9 \text{ nm}$ (green), $d = 8 \text{ nm}$ (orange) and $d = 6 \text{ nm}$ (dark yellow) and similar composition, as extracted from the external magnetic field measurement (see Fig.~1). The black dashed line represents the flow diagram of the iQHE and the red dashed line two parallel conducting topological surface states.}%
\label{fig:ScalingDiffd}%
\end{figure}

Finally, we remark on the significance of the protective Te cap. Samples which are grown without a cap and otherwise processed in the same way, also exhibit the QAHE effect. The flow diagrams extracted from the external magnetic field dependence of three uncapped layers are shown in supplementary Fig.~6. All three of these layers are at least 9 nm thick as determined by XRR measurements, yet show 2D scaling behavior. The flow diagram of uncapped layers therefore matches the scaling behavior of thinner capped layers. This is likely to be a result of degradation in the topmost layers due to exposure of the unprotected surface to ambient conditions and the lithography process, and suggest that uncapped samples have a dead layer at their surface, and thus are effectively thinner than their nominal values. 

The range of thickness at which we see the scaling transition in capped layers is not inconsistent with the expected opening of the hybridization gap at ~6 nm. While most of the samples reported on in literature are below this threshold, a clear iQHE scaling behavior is also seen in measurements from Checkelsky \textit{et al.}~\cite{Checkelsky2014} and Bestwick \textit{et al.}~\cite{Bestwick2015} on layers which are nominally 8 nm and 10 nm thick, respectively. Considering that magnetically doped (Bi,Sb)$_2$Te$_3$ layers have a considerable roughness it may be that the percolation path of the edge channel through the sample contains thinner parts which could explain an onset of this behavior in nominally thicker samples. Another explanation could be that the various capping methods (a 2 nm thin Al layer grown \textit{in situ}~\cite{Bestwick2015} or  \textit{ex situ} grown AlO$_x$ ~\cite{Checkelsky2014}) have a different degree of effectiveness than our Te cap. 

{\it Speculation relating to the direction of magnetization in our samples.} A fine point which should be commented on relates to the direction of the magnetization vector in the sample. In the geometry examined in the theoretical literature the magnetization is usually made to point out of the sample on all surfaces. Whether this is a necessity or how the axion physics would look for fields going inwards and outwards on opposite surfaces is not yet explored. Moreover, no direct measurement of the magnetization has yet been reported on at low enough temperatures for the samples to be in the quantized regime, where the character of the magnetization reversal is qualitatively different than that at higher temperature (See Fig.~S10 in \cite{Bestwick2015} for example). It is not clear if in our samples the magnetization is homogeneous, or if energy considerations (discussed in the supplementary material, which includes Ref. to \cite{Nunez2012}) lead to an always inward or always outward configuration. In either event, while more investigation into this issue is certainly needed before clarity is achieved, the very fact that we observe the quantum anomalous Hall effect in the samples proves that the edge states survive, and that their scaling behavior can be analyzed.   

A single species Dirac fermion on a 3D TI surface state is linked to axion electrodynamics based on gauge symmetry arguments. Scaling analysis performed on our layers is consistent with that expected from such an axion system. While uncertainty about the magnetization state leave some degree of speculation in associating our scaling to an axion insulator, we suggest that it is a plausible explanation based on currently available evidence.

{\it Conclusion.} We have studied the flow diagram of the QAHE in several layers. From the scaling behavior we determine that the QAHE in our capped 9 nm thick V-doped (Bi,Sb)$_2$Te$_3$ layers originates from the two topological surface states of the magnetic 3D TI each contributing the half-integer quantization of $\sigma_{xy}=\frac{1}{2}\text{ e}^2/\text{h}$ to the total Hall conductivity. The center of the semicircle in the flow diagram being shifted from a finite value to zero is robust evidence, which does not rely on discriminating between the various surface contributions. This result qualitatively differs from most previous publications showing 2D behavior~\cite{footnote1}
and is a robust transport observation of a distinct QAHE having scaling properties consistent with one resulting from the presence of axionic action characterizing the electrodynamic response of a magnetic 3D TI, i.e., an axion insulator.  

\begin{acknowledgments}
We gratefully acknowledge the financial support of the EU ERC-AG Program (project 3-TOP), the EU ERC-StG Program (project TOPOLECTRICS), the DFG through SFB 1170 ``ToCoTronics'' and the Leibniz Program.
\end{acknowledgments}

{\it Note added.}-In addition to note~\cite{footnote1}, an additional publication appeared~\cite{Mogi2017}, during the review process of our manuscript, in which a 3D-like scaling (as the red dashed line of our Fig.~2) is reported in Figure S6 of their supplementary material. However, as in~\cite{footnote1} the significance of this observation was not recognized. Interestingly, in the configuration where the authors believe to have achieved a magnetization pointing out of the plane on both surfaces, the observed scaling is that of our 2D geometry (as the black dashed line of our Fig.~2), which does not support axionic responce.

\setcounter{figure}{0} 

\bigbreak
\begin{center} 
\large
\textbf{Supplementary Material }
\normalsize
\end{center}
\bigbreak

{\it Sample preparation.} After MBE growth, the samples are processed by standard optical lithography techniques. In a first lithographic step, a six-terminal Hall bar geometry mesa is defined using Ar ion beam etching, with positive tone resist as an etching mask. In a second lithographic step, after Ar ion beam etching removal of the tellurium cap from selected parts of the defined mesa, the sample is transferred in-situ under vacuum conditions, into metalization chamber to evaporate the AuGe contact leads. A top gate oxide, consisting of 20 nm of AlOx, and 1 nm of HfOx, is deposited on the structure using atomic layer deposition, and immediately covered with a Au gate electrode in the third lithographic step. The remaining insulating oxide layer is removed from the leads area with water diluted HF in the final step. Afterwards, each processed sample is glued to a chip carrier, and an ultrasonic bonder is used to connect Au wires to the device. Each sample consists of two Hall bar devices, big and small, with widths of 200 $\mu$m and 10 $\mu$m, and separations of 600 $\mu$m and 30 $\mu$m, respectively, between adjacent contacts. No difference in transport is observed between the two. Supplementary Fig.~1 shows an optical microscope photograph of a typical sample.

{\it Transport data for layers with thicknesses beyond $d$ = 9 nm.} Supplementary figure 2 presents longitudinal (a) and Hall (b) magneto-resistivity data for V$_{0.1}$(Bi$_{0.21}$Sb$_{0.79}$)$_{1.9}$Te$_3$ layers with thicknesses $d$ = 9 nm, $d$ = 15 nm, and $d$ = 25 nm grown under the same conditions. The layer with thickness $d$ = 9 nm is a representative one exhibiting perfect quantization \cite{Grauer2015}, plotted here as a comparison to the thicker layers. The layer with $d$ = 15 nm has no Te protecting cap, whereas both with $d$ = 9 nm and $d$ = 25 nm have one. Measurements were performed at base temperature (nominally 25 mK) of the dilution refrigerator, and gate voltage providing the biggest Hall resistivity for each of the samples. Layer thicknesses were obtained from  X-ray reflection (XRR) measurements. For the thicker samples, the reduced values of the anomalous Hall signal and increased four terminal longitudinal resistance are a reflection of the fact that the bulk of the layer begins to contribute parasitic conduction. As such, it is impossible to discriminate between the bulk and topological surface contributions to the conductivity, and thus impossible to examine the scaling behavior of the surface state transport.

\begin{figure}[b] 
\includegraphics[width=\columnwidth]{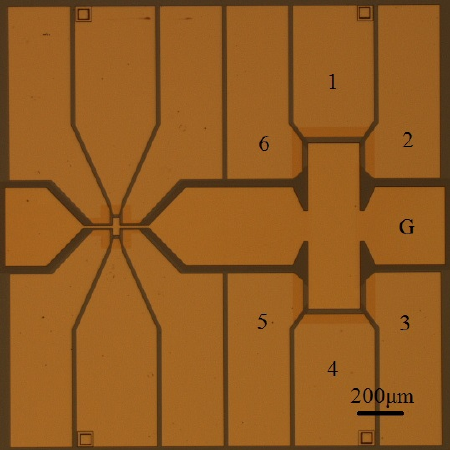}%
\caption {Optical microscope photograph of one of the processed samples consisting of two Hall bar devices, big (600 $\mu$m x 200 $\mu$m) and small (30 $\mu$m x 10 $\mu$m). Contacts (1) and (4) are used to pass a current through the device, contacts (2)(3)(5) and (6) are used for proper 4-terminal measurements, and contact (G) is used for applying a gate bias.}
\label{Fig:FigS1}
\end{figure}

\begin{figure}[] 
\includegraphics[width=\columnwidth]{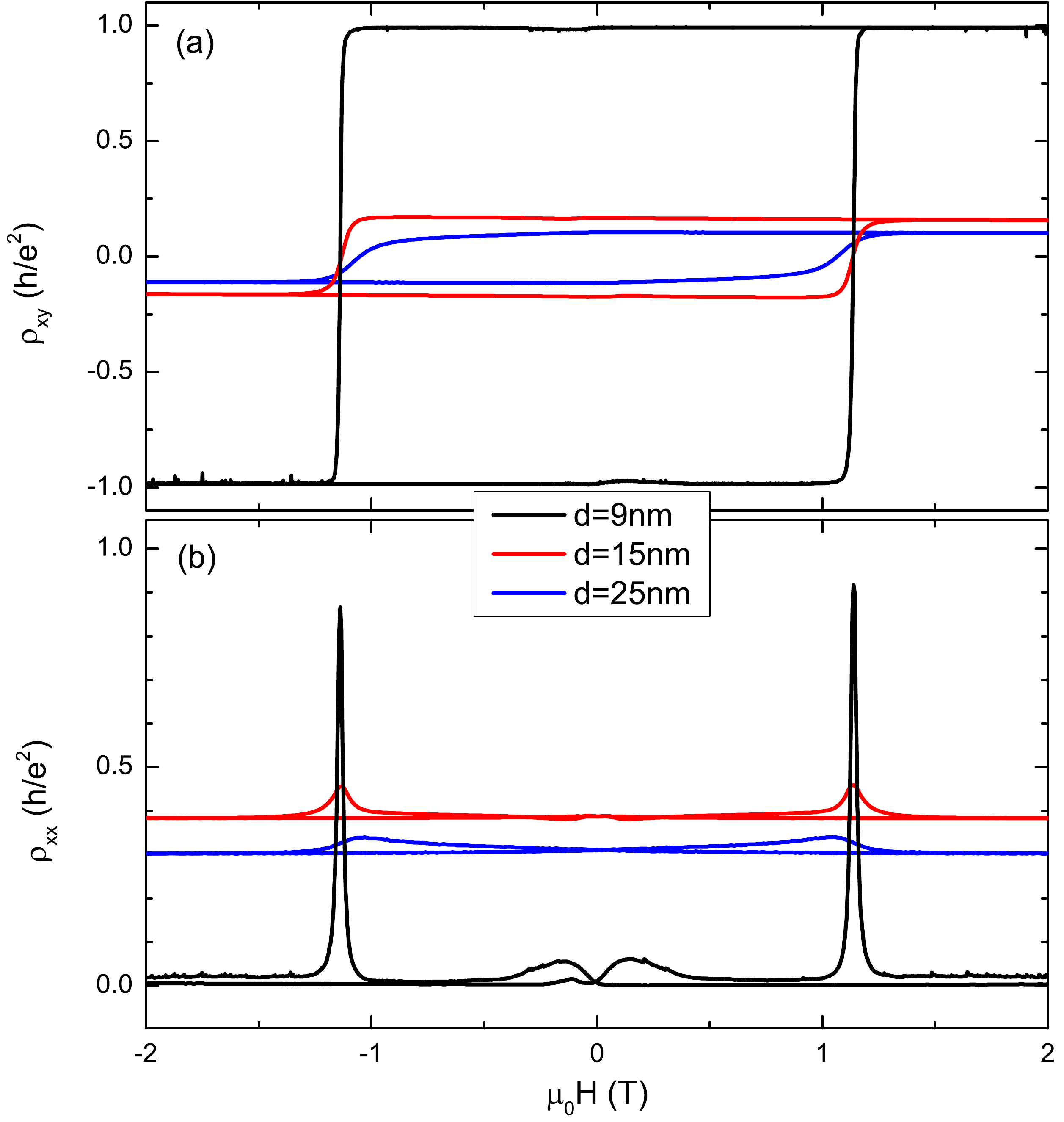}%
\caption {The Hall (a) and longitudinal (b) magneto-resistivity data for $d$ = 9 nm (black), $d$ = 15 nm (red), and $d$ = 25 nm (blue) thick V$_{0.1}$(Bi$_{0.21}$Sb$_{0.79}$)$_{1.9}$Te$_3$ samples grown under the same conditions. The sample with $d$ = 15 nm has no protective cap, while the other two were grown with Te cap. Measurements were performed at base temperature of the dilution refrigerator for gate voltage providing maximal Hall resistivity.}
\label{fig:FigS4}
\end{figure}

\begin{figure} 
\includegraphics[width=\columnwidth]{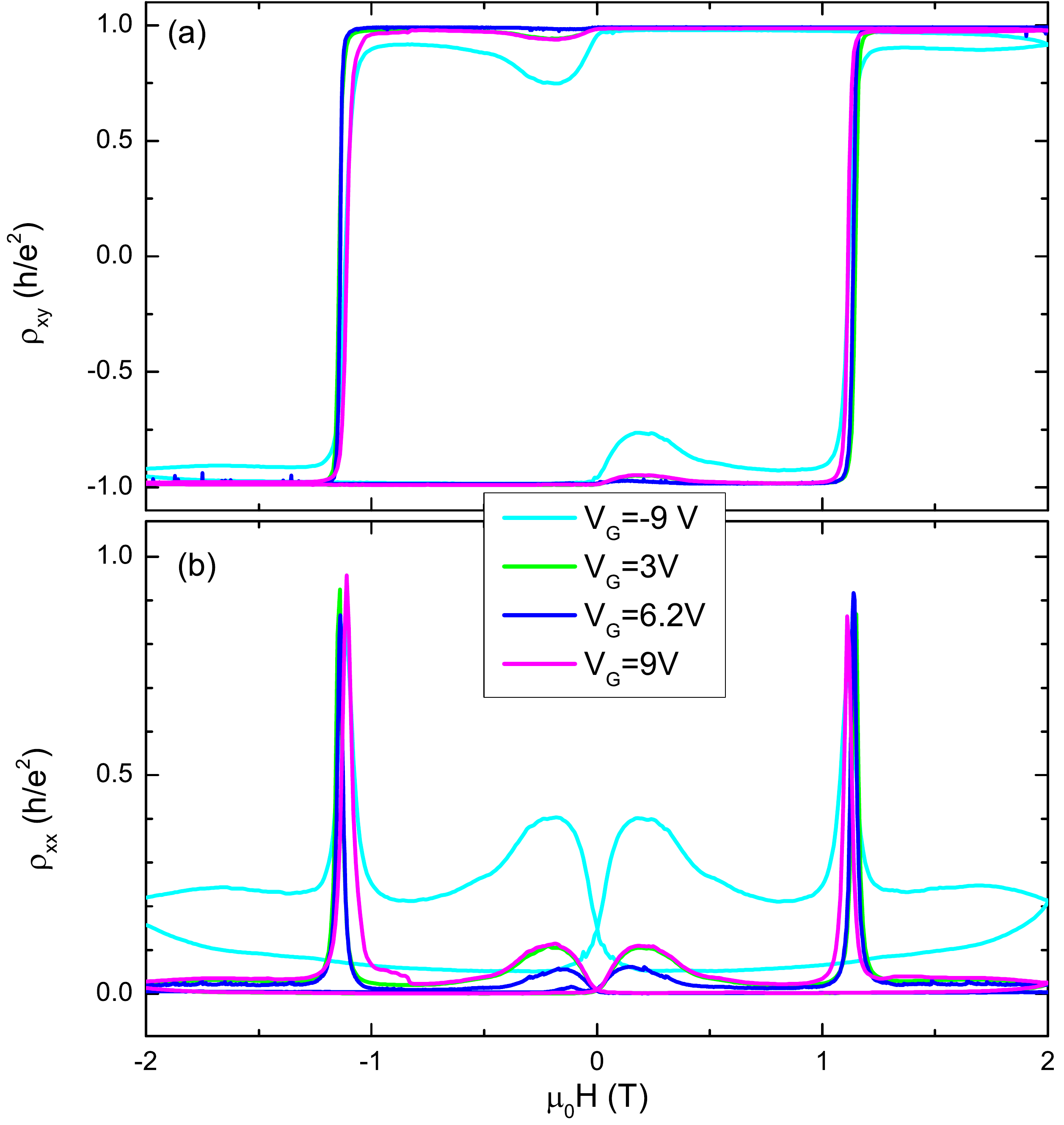}%
\caption {The Hall (a) and longitudinal (b) magneto-resistivity data for four fixed gate voltage values $V_G$ = -9 V (cyan), $V_G$ = 3 V (green), $V_G$ = 6.2 V (blue), and $V_G$ = 9 V (magenta), for a representative 9 nm thick V$_{0.1}$(Bi$_{0.21}$Sb$_{0.79}$)$_{1.9}$Te$_3$ layer exhibiting perfectly quantized transport (Fig. 3 in the letter, same color coding). Measurements were performed at base temperature of the dilution refrigerator.}
\label{fig:FigS6}
\end{figure}

\begin{figure} 
\includegraphics[width=\columnwidth]{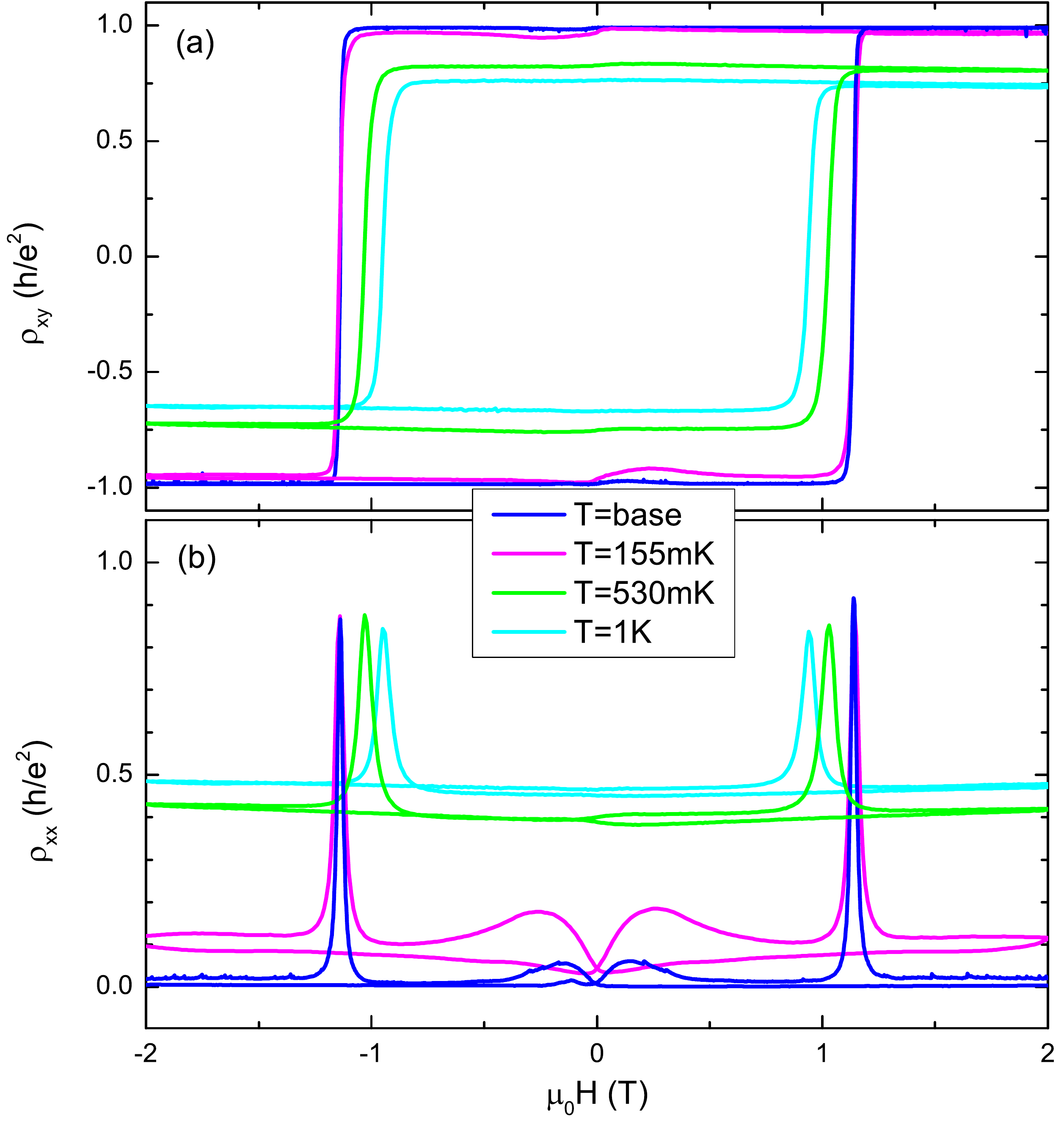}%
\caption {The Hall (a) and longitudinal (b) magneto-resistivity data for four fixed temperatures $T$ = 155 mK (magenta), $T$ = 530 mK (green), $T$ = 1 K (cyan), and base temperature of the dilution refrigerator (blue), for a representative 9 nm thick V$_{0.1}$(Bi$_{0.21}$Sb$_{0.79}$)$_{1.9}$Te$_3$ layer exhibiting perfectly quantized transport (Fig. 4 in the letter, same color coding). Measurements were performed at base temperature of the dilution refrigerator.}
\label{fig:FigS8}
\end{figure}

\begin{figure} 
\includegraphics[width=\columnwidth]{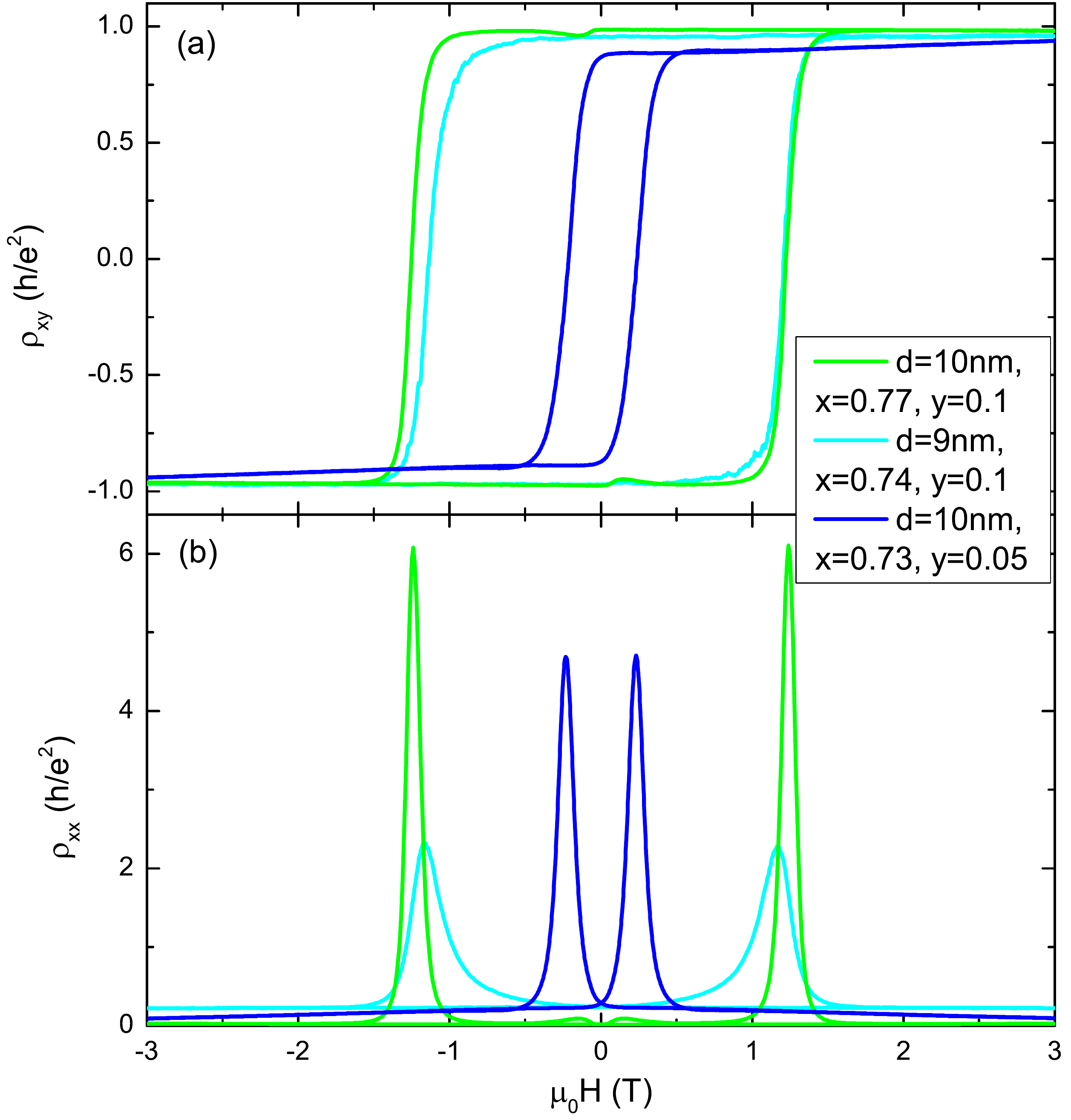}%
\caption {Hall (a) and longitudinal (b) magneto-resistivity data for three V$_{y}$(Bi$_{1-x}$Sb$_{x}$)$_{2-y}$Te$_3$ uncapped layers (scaling plots acquired from this data are presented in supplementary Fig.~6,  with the same color coding). The composition and thickness of the layers are V$_{0.1}$(Bi$_{0.23}$Sb$_{0.77}$)$_{1.9}$Te$_3$ and $d = 10 \text{ nm}$ (green), V$_{0.1}$(Bi$_{0.26}$Sb$_{0.74}$)$_{1.9}$Te$_3$ and $d = 9 \text{ nm}$ (cyan), V$_{0.05}$(Bi$_{0.27}$Sb$_{0.73}$)$_{1.95}$Te$_3$ and $d = 10 \text{ nm}$ (blue).}
\label{fig:FigS7}
\end{figure}

\begin{figure} 
\includegraphics[width=\columnwidth]{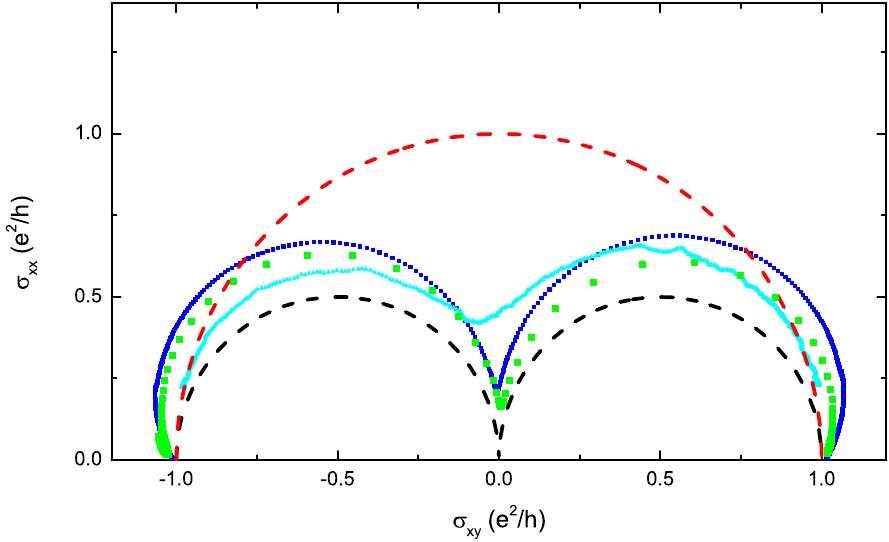}%
\caption{Flow diagram mapping $\sigma_{xx}$ to $\sigma_{xy}$ for 3 uncapped layers grown on Si(111). The composition and thickness of the layers are V$_{0.1}$(Bi$_{0.23}$Sb$_{0.77}$)$_{1.9}$Te$_3$ and $d = 10 \text{ nm}$ (green), V$_{0.1}$(Bi$_{0.26}$Sb$_{0.74}$)$_{1.9}$Te$_3$ and $d = 9 \text{ nm}$ (cyan), V$_{0.05}$(Bi$_{0.27}$Sb$_{0.73}$)$_{1.95}$Te$_3$ and $d = 10 \text{ nm}$ (blue). The black dashed line represents the flow diagram of the iQHE and the red dashed line two parallel conducting topological surface states. }%
\label{fig:ScalingUncapped}%
\end{figure}

\begin{figure} 
\includegraphics[width=\columnwidth]{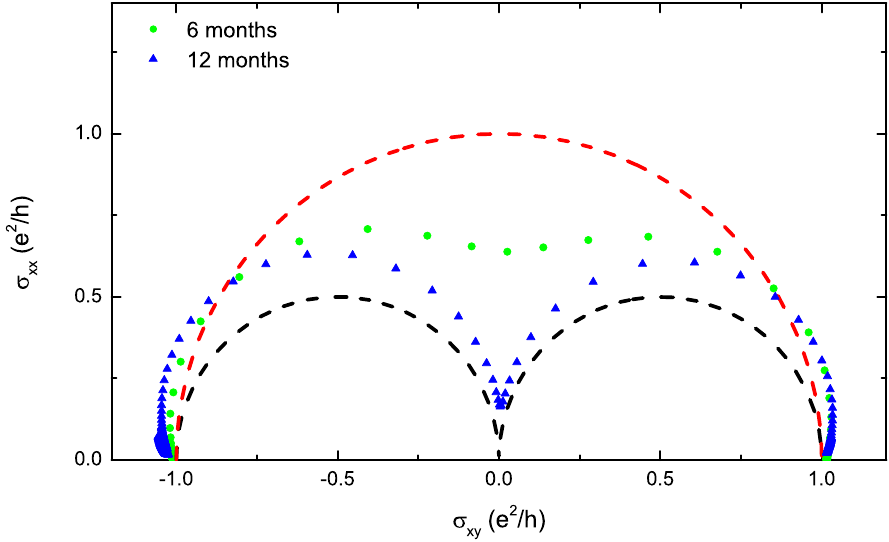}%
\caption {Flow diagram mapping $\sigma_{xx}$ to $\sigma_{xy}$ using external magnetic field as driving parameter, collected at the base temperature of the dilution refrigerator, for the gate voltage providing maximal Hall resistivity for the two 10 nm thick V$_{0.1}$(Bi$_{0.23}$Sb$_{0.77}$)$_{1.9}$Te$_3$ samples without protective Te cap, processed from the same grown layer. In the case of the first sample, the surface was exposed to ambient conditions for 6 months before processing, and in case of the second one, for 12 months. (Scaling plot for the sample exposed for 12 months is visible in supplementary Fig.6 as well)}
\label{fig:FigS5}
\end{figure}

\begin{figure} 
\includegraphics[width=\columnwidth]{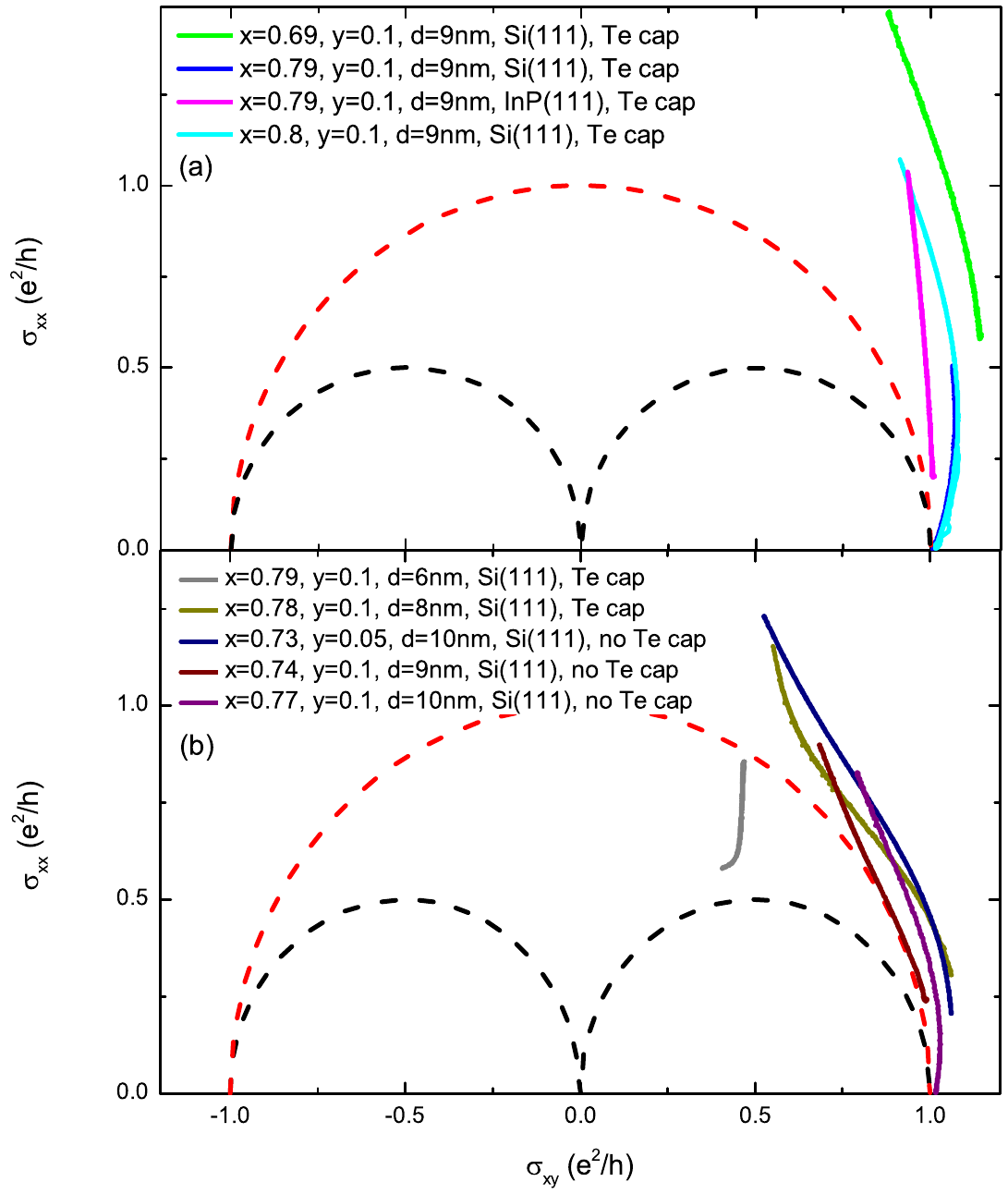}%
\caption {Flow diagrams mapping $\sigma_{xx}$ to $\sigma_{xy}$ with temperature as a driving parameter, collected during cooling from $T$ = 1 K to base temperature of the dilution refrigerator, at the gate voltage providing maximal Hall resistivity for each, and no external magnetic field applied. The measurements are preceded by magnetic field treatment to saturate the magnetization of the samples. Scaling of various V$_{y}$(Bi$_{1-x}$Sb$_{x}$)$_{2-y}$Te$_3$ layers which are (a) in the 3D regime and (b) in the 2D and intermediate regimes are presented. The black dashed line represents the flow diagram of the iQHE and the red dashed line two parallel conducting topological surface states.}
\label{fig:FigS2}
\end{figure}

\begin{figure} 
\includegraphics[width=\columnwidth]{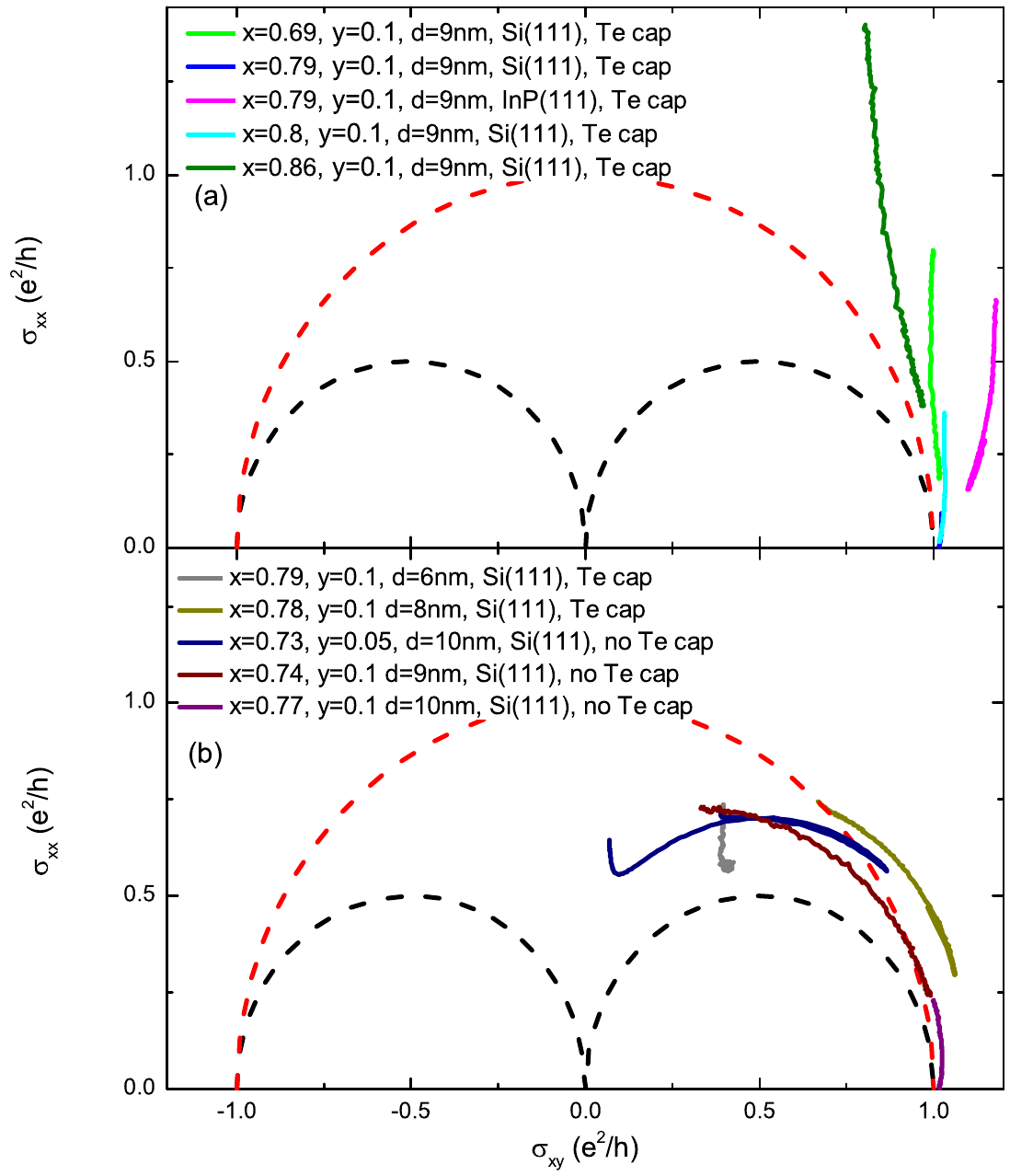}%
\caption {Flow diagrams mapping $\sigma_{xx}$ to $\sigma_{xy}$ with gate voltage as driving parameter, collected at base temperature of the dilution refrigerator and no external magnetic field applied, after a magnetic field treatment to saturate magnetization of the layers. Results for various V$_{y}$(Bi$_{1-x}$Sb$_{x}$)$_{2-y}$Te$_3$ layers from the letter in the 3D regime, Te capped with thickness $d$ $\approx$ 9 nm (a), and samples in the 2D and intermediate regimes (b) are presented (including uncapped thick samples exhibiting behavior of effectively thinner layers). The black dashed line represents the flow diagram of the iQHE and the red dashed line two parallel conducting topological surface states.}
\label{fig:FigS3}
\end{figure}

{\it Magneto-resistivity data for scaling plots presented in Fig.~3 and Fig.~4 of the Letter.} Supplementary Fig.~3 presents raw magneto-resistivity data for the scaling plots in Fig.~3 of the Letter. Magnetic field scans were performed at base temperature of the dilution refrigerator for various fixed gate voltage values, ranging from $V_G$ = -9 V to $V_G$ = 9 V (dielectric break limit of the 21 nm thick gate insulator on our samples). 

Supplementary Fig.~4 shows raw magneto-resistivity data for scaling plots from Fig.~4 in the Letter. Magnetic field measurements were conducted at various fixed temperature values ranging from base temperature of the dilution refrigerator, up to $T$ = 1 K, at the gate voltage providing the highest anomalous Hall response.

{\it Aging of the uncapped samples.} Supplementary Fig.~6 presents scaling analysis with magnetic field as the driving parameter, for three layers grown without protective Te cap. Magneto-resistivity data from which the scaling plots were acquired are shown in Supplementary Fig.~5. Measurements were performed at base temperature of the dilution refrigerator, for the gate voltage providing the biggest anomalous Hall resistivity for each sample. All three layers are at least 9 nm thick as determined by XRR measurements, yet show 2D scaling behavior. The flow diagram of uncapped layers therefore matches the scaling behavior of thinner capped layers. Supplementary Fig.~7 shows similar scaling analysis for two samples processed from one of the uncapped 10 nm thick layers presented in supplementary Fig.~6 (V$_{0.1}$(Bi$_{0.23}$Sb$_{0.77}$)$_{1.9}$Te$_3$). One of them was exposed for 6 months to ambient conditions, and the other for 12 months between growth and processing. The flow diagram reveals an evolution of the scaling behavior towards the iQHE one, with increasing layer exposure time. This strongly indicates that samples without a proper protective cap exhibit behavior analogous to thinner grown capped layers due to degradation of the top atomic layers.

{\it Magnetic field as a parameter for quantum anomalous Hall effect flow diagram investigation.} As mentioned in the main text, any parameter which causes the sample to transition between plateaus can be used as the drive parameter to explore the scaling, and various ones have been used over the years, including gate voltage, temperature, and external magnetic field. In this letter we focused our attention on using magnetic field as a parameter for QAHE scaling analysis, because it is the parameter that reveals the largest part of the scaling diagram, and the one which minimizes any possible influence of parasitic bulk effects. For completeness, we have analyzed the temperature and gate voltage driven scaling dependence as well (Supplementary Fig.~8 and supplementary Fig.~9).  

{\it Scaling analysis with temperature as a parameter.} Supplementary Fig.~8 shows a scaling analysis of the samples from the letter with temperature as the driving parameter, collected during cooling from $T$ = 1 K to base temperature of the dilution refrigerator, at gate voltages providing maximal Hall resistivity for each, and no external magnetic field applied, after a magnetic field treatment to saturate the magnetization of the layers. The graph includes analysis of samples in a 3D regime with thickness $d$ $\approx$ 9 nm (a), and samples in intermediate and 2D regimes (including thick uncapped layers exhibiting behavior of thinner layers) (b). There are indications of a difference in the flow diagram between the two situations, however the sampled parameter space is too small to draw any strong conclusions regarding presence of the axionic screening. Moreover, we caution that the matrix inversion formalism used to calculate conductivity tensor elements is valid only in a 2D system, or in case of a 3D topological insulator (TI) slab, parallel 2D topological surface states separated by insulating bulk. For this reason, it might not be possible to reliably extract $\sigma_{xx}$ and $\sigma_{xy}$  for elevated temperatures in a 3D TI, due to the likely appearance of trivial and temperature dependent conduction in the bulk. This raises questions as to the reliability of analysis based on temperature driven scaling behavior in such systems. 

{\it Scaling analysis with gate voltage as a parameter.} The gate voltage parameter dependence of the flow diagrams presented in supplementary figure 9 exhibits a clear difference between scaling of the layers in the 2D and intermediate regimes (including uncapped thick samples exhibiting behavior of thinner layers) (b), and those in 3D regime with thickness $d$ $\approx$ 9 nm (a). The measurements were performed at base temperature, and no external magnetic field applied, preceded by magnetic field treatment to saturate magnetization of the samples. The former appear to follow iQHE scaling, while the latter reveal different behavior. Similarly to temperature dependence, the range of revealed scaling flow is significantly smaller than for magnetic field measurements. Nevertheless, the flow diagrams visible in (a) and (b) clearly show distinct behavior, fully consistent with that observed in the magnetic field driven measurements.

{\it Energy considerations for the magnetization state.} As we commented on in the main text, there has yet to be any direct experimental measurements of the magnetization state in samples in the quantum anomalous Hall regime, but there is significant evidence that the state is rather complex. This includes the reports on partially superparamagnetic like behavior in Ref. \cite{Grauer2015}, as well as the clearly distinct qualitative magnetization behavior, below and above approximately 350 mK, as seen in Fig.~S10 in the supplemental to Ref. \cite{Bestwick2015}. This may all suggest that the ferromagnetic transition observed at $T_c$ = 23 K \cite{Chang2015a} is that of the bulk material, whereas the magnetic character of the surface state itself has some independent character.     

To this last point we would note that the authors of Ref. \cite{Nomura2011} raise the issue of domains with opposite orientation of magnetisation on a single surface, and how it can be discarded on energy principles. This issue requires analyzing the balance of three different energy scales: (i) single-ion magnetic anisotropy (see also \cite{Nunez2012}) (ii) Zeeman energy of the magnetic ions relating to the effectively ferromagnetic coupling between the ions and (iii) the magneto-electric energy itself, which is Eq. 3 in Ref. \cite{Nomura2011}. As one compares all three energy scales, following the line of thought in Ref. \cite{Nomura2011}, it seems likely that (iii) wins beyond a sufficient electric field strength. This is plausible as the other processes do not benefit from the electric field strength present at the thin film surfaces, and all energies (i)-(iii) scale linearly as a function of magnetic field, i.e. as a function of magnetic impurity density. 

Having established that oppositely oriented magnetic impurity domains are not preferred for energetic reasons on a single surface, as done in Ref. \cite{Nomura2011}, it is conceivable that maximising the magneto-electric energy gain from the surface magnetic ion composition is likewise the driver for making all magnetic impurities point in or out of the TI. Judging from the estimate of Ref. \cite{Nomura2011}, the magneto-electric term is dominant over e.g. the Zeeman energy by possibly one or even more orders of magnitude. For the very same reasoning, one could expect the magnetic ion polarization to point all out-of-surface or in-surface. 
Exploring this hypothesis beyond the above energy scale argument would certainly be interesting, but would be very challenging from the simulation point of view, and well beyond the scope of this manuscript. It would require analyzing microscopically a 3D system of disordered magnetic ions doped into a topological insulator, which appears to be out of reach of current models.


%

\end{document}